\documentclass[11pt]{article}
\usepackage{graphicx,citesort,ils08eng,amssymb,amsmath}

\begin{document}

\begin{center}

{\LARGE{\bf Hubble law, Accelerating Universe and Pioneer Anomaly as
effects of the space-time conformal geometry}}

\vskip7.5mm

{\large L.M.~Tomilchik}

\vskip5mm

\textit{B. I. Stepanov Institute of Physics, NAS of Belarus, Minsk, Belarus \\ 
lmt@dragon.bas-net.by}

\vskip10mm

{\bf Abstract}

\vskip2.5mm

\parbox[t]{110mm}{{\small
The description of the cosmological expansion and its possible
local manifestations via treating the proper conformal
transformations as a coordinate transformation from a comoving
Lorentz reference frame to an uniformly accelerated one is
given. The explicit form of the conformal time inhomogeneity is
established. The  expression defining the location cosmological
distance in the form of simple function on the red shift is
obtained. By coupling it with the relativistic formula for the 
longitudinal Doppler effect, the explicit expression for the Hubble law 
is obtained, which gives rise to the connection between acceleration
and the Hubble constant. The expression generalizing the conventional
Hubble law reproduces kinematically the experimentally observed
phenomenon treated conventionally as a Dark Energy manifestation.
The conformal time deformation in the small time
limit leads to the quadratic time nonlinearity. Being applied to
describe the location-type experiments, this predicts the
existence of the universal uniformly changing blue-shifted frequency drift.
The obtained formulae reproduce the Pioneer Anomaly experimental data.}}

\end{center}

\vskip5mm

\section*{Introduction}

About ten years ago the primary communications regarding two significant
astrophysics discoveries were appeared. Two independent research groups announced
in 1998--1999 \cite{1,2} about the detection of the peculiar deviations from the Hubble
law linearity near the cosmological red shift value equal to $z_{exp}=0.46\pm0.13$.
At 1998 the information was published \cite{3} (see also \cite{4}) about the reliable
experimental registration of the systematic uniform anomalous blue frequency drift
in the radiosignals received from the spacecrafts Pioner 10/11 leaving the Solar
system (Pioneer Anomaly --- PA).

The standard treation of the first of these phenomena leads to a predominance (more than 65 \% from the total energy amount) of the paradoxical dark energy in contemporary Methagalaxy. The conventional treation of the observable PA-effect as a common Doppler shift is nessesarily presumed the existence of some additional uniform sunward acceleration having the magintude equal to $a_p=(8.74\pm1.33)\cdot10^{-8}\ 
\frac{\mbox{\small cm}}{\mbox{\small s}}$ experienced by the spacecrafts. In spite of the abundant theoretical suppositions concerning the possible source of such an acceleration the origin of the Pioneer Anomaly remains unexplained up to now \cite{5}.

The proximity of $a_p$ to the quantity $W_0=cH_0$ ($c$ is the speed of light, $H_0$ is the Hubble
constant) was noted by several authors. However, any unambiguous theoretical
arguments in favor of the possible existence of such a connection
are absent up to now.

In the present report the description of the cosmological expansion
and its possible local manifestations is given via treating the
proper conformal transformations as a coordinate transformation
from a comoving Lorentz RF to an uniformly accelerated RF. Such an
approach permit to derive the explicit analytic expression for
the Hubble law, which allows to connect the acceleration
with the Hubble constant as well as to reproduce the
Accelerating Universe and Pioneer Anomaly effects in exact correspondence with the
observations. These effects, once dictated by the conformal time
inhomogeneity, can be interpreted as the manifestations
of the background acceleration existence, i.e., of
the noninertial character of any physical frame of reference which
is coupled locally to an arbitrary point of the modern Metagalaxy.

The topics of the present report are as follows.
\begin{itemize}
\item[I.] Conformal transformations and the accelerating frame of reference.
\item[II.] Conformal deformation of the light cone and time inhomogeneity.
\item[III.] The dependence of distance on the red shift.
\item[IV.] The Hubble law. Connection between the Hubble constant and the
background acceleration.
\item[V.] Accelerating Universe effect without Dark Energy.
\item[VI.] The universal blue-shifted frequency drift. Pioneer Anomaly.
\end{itemize}

\section*{I. Conformal transformations and the accelerating frame of reference}
It is well known that the special conformal transformations (SCT)
$$x'^\mu = \frac{x^\mu +a^\mu(x^\alpha
x_\alpha)}{1+2(a^\alpha x_\alpha)+(a^\alpha a_\alpha) (x^\beta
x_\beta)}, \eqno(1)$$ where $a^\mu$ is the four-vector parameter,
$\eta_{\mu\nu}=\mathrm{diag}(1,-1,-1,-1)$, can be interpreted as
the transformations between the Lorentz (comoving) frame of
reference $S(x^\mu)$ and the noninertial (accelerated) frame of
reference $S'(x'^\mu)$ (see, for example, \cite{6}). Following \cite{6} we
shall for simplicity consider a two-dimensional subspace
$\{t,x\}$, i.e. we put $$x^\mu=(ct,x,0,0),\
a^\mu=(0,-\frac{w}{2c^2},0,0),\eqno{(2)}$$ where $w$ is a
constant uniform acceleration. In general $a^\mu$ is related with a constant
4-acceleration $w^\mu$ as follows $a^\mu=w^\mu/2c^2$.

It is conveniently to write the transformations (1) in the following noncovariant
form: $$\left.\begin{gathered}x'=
\frac{x\xi-\frac{wt^2}{2}}{\xi^2-\eta^2},\\
t'=\frac{t}{\xi^2-\eta^2},
\end{gathered}\right\} \eqno{(3)}$$
where $$\xi=1+\frac{wx}{2c^2},\qquad \eta=\frac{wt}{2c},\qquad r_0=c^2/w.\eqno(4)$$

In the case when $\frac{wx}{2c^2}$ and $\frac{wt}{2c}$ are negligible we have from (3) $$x'=x-\frac{wt^2}{2},\qquad t'=t,\eqno{(4)}$$ which corresponds to Galilei-Newton kinematics. It is also clear from (3) that sign $(-)$ of the vector's $a^\mu$ $x$-component describes a positive direction of $S'$ acceleration along the $x$-axis of the inertial reference frame (IRF) $S$.

The velocity and acceleration transformations laws are presented in \cite{7}. In the approximation $\xi\approx1,\ \eta\ll1$ this transformation leads to formulae $$V'_x=V_x-wt,\eqno{(5)}$$ for the velocities and to $$w'_x=w_x-w\eqno(6)$$ for the accelerations in the correspondence with Galilei-Newton kinematics.

From (5), the existence of blue Doppler drift follows immediately. Really, due to (5) the longitudinal component of a point velocity $V'_x$ measured by an observer fixed in non-inertial RF $S'$ will be less then that measured by an observer fixed in the inertial comoving RF $S$ by $\Delta V_x=wt$.

Therefore, for the observed blue shift we have
$$\Delta\nu_{obs}=\nu'-\nu_{mod}=\nu_0\frac{wt}{c},\eqno{(7)},$$
where $\nu_0$ is the signal frequency emitted by a source fixed in $S$
and $\nu_{mod}$ is the frequency defined by neglecting
non-inertiality of $S'$. In the approximation considered the shift
is linear in time. The rate of shift $\dot\nu_{obs}=
\frac{d\nu_{obs}}{dt}$ is defined by the following simple relation
$$\dot\nu_{obs}=\nu_0\frac{w}{c}.\eqno{(8)}$$

This result, in principle, is well known. Here we have to do, in
fact, with the effect of the gravitational (Einstein) frequency
drift described on the basis of the equivalence principle.

The relation (6) holds in every comoving RF as long as in this frame $\xi\approx1,\ \eta\ll1$ approximation
is valid. A probe particle free of dynamical influence in the comoving RF, i.e. with $W_x=0$, in accordance with (5) will be uniformly accelerated in the non-inertial RF $S'$ with the
constant acceleration of $-w$. Such an acceleration can be registered by any observer fixed in any point of this non-inertial RF $S'$. By the equivalence principle the non-inertial observer is
entitled to identify this acceleration with an existence of a constant (background) gravitational field which results in acceleration $w$.

This result can be confirmed via dynamical approach using the expression
of the conformal one-particle Lagrangian as it shown in \cite{8}.
Following a common practice for the Lagrangian dynamical description in the
noninertial RF the following expression for the one-particle
nonrelativistic Lagrangian can be obtained $$L_{NR}=\frac{m_0\vec v^2}{2}-\frac{m_0c^2}{R_u}r.\eqno{(9)}$$
where $$R_u=r_0=\frac{c^2}w.$$
The term $\frac{m_0c^2}{R_u}r$ in (9) can be treated in the correspondence with the equivalence principle
as a potential energy of the probe mass $m_0$ in the following ''gravitational'' potential:
$$\varphi_0=\frac{c^2}{R_u}r=wr.\eqno{(10)}$$

This potential plays the role of the source of the background uniform acceleration having the magnitude
$w=w_{background}$ and directed towards the point of observation.

\section*{II. Conformal deformation of the light cone and time inhomogeneity}
Now we consider the transformations of the light cone generatrices
under SCT (1). Since $x'^\mu x'_\mu=x^\mu x_\mu(1+2(ax) +
(a)^2(x)^2) ^{-1}$, transformations (1) leave the light cone
equation invariant, i.e., from $x^\mu x_\mu=0$ follows $x'^\mu
x'_\mu=0$. However the light cone surface is deformed
non-linearly. From (1), it generally follows
$$x'^\mu=\frac{x^\mu}{1+2(ax)},\eqno{(11)}$$ when additionally
$x^\mu x_\mu=x'^\mu x'_\mu=0$.

In the two-dimensional case considered we have the relation
$$t'_\pm=\frac{t}{1\pm\frac{t}{t_{lim}}},\eqno{(12)}$$ where
$t_{lim}=c/w$.

The choice of sign corresponds to signal propagation in the
forward and backward directions, respectively. We are reminded
that the symbol $t$ ($t'$) represents time of a light signal
propagation between two spatially separated points in the space of
Lorentz RF $S$ (in the non-inertial RF $S'$). So the quantities
$R=ct$ and $R'=ct'$ define location distances in both of these
RFs.

Obviously a semi-infinite time interval $0\leqslant t<\infty$
corresponding to the positive (forward) direction of signal
propagation maps onto a finite time interval $0\leqslant t'_+
\leqslant t_{lim}$. For the backward direction, on the contrary, a
finite interval $0\leqslant t \leqslant t_{lim}$ maps onto a
semi-infinite time interval $0\leqslant t'_-<\infty$ (Fig. 1).

\begin{figure}[h!]
\centering \includegraphics[width=0.48\textwidth]{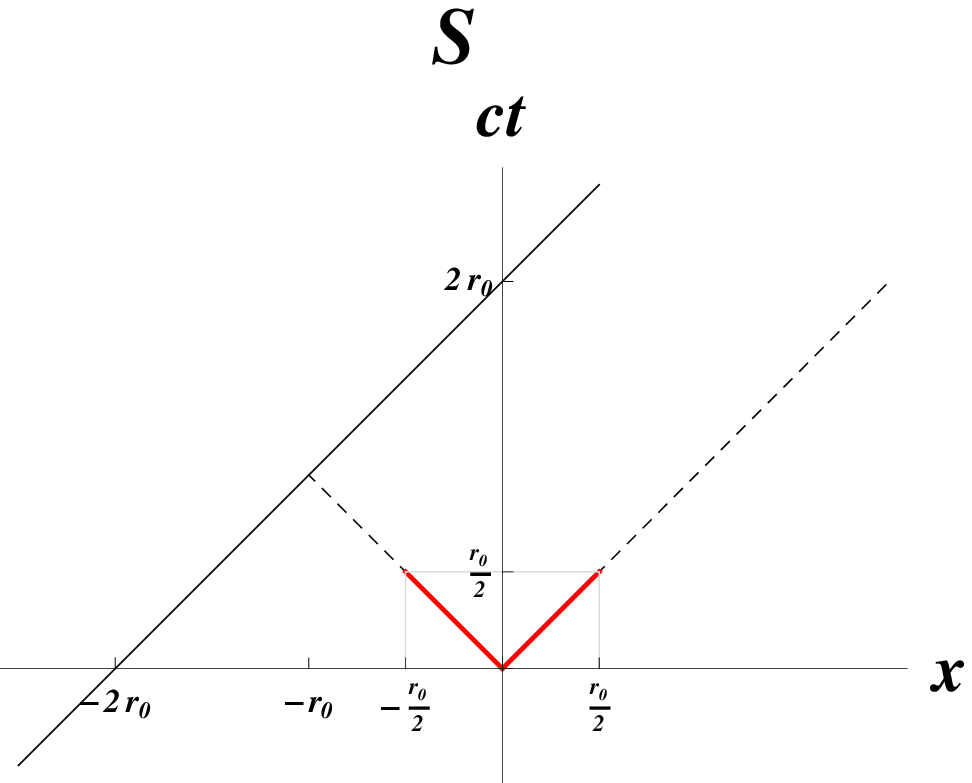}
\includegraphics[width=0.48\textwidth]{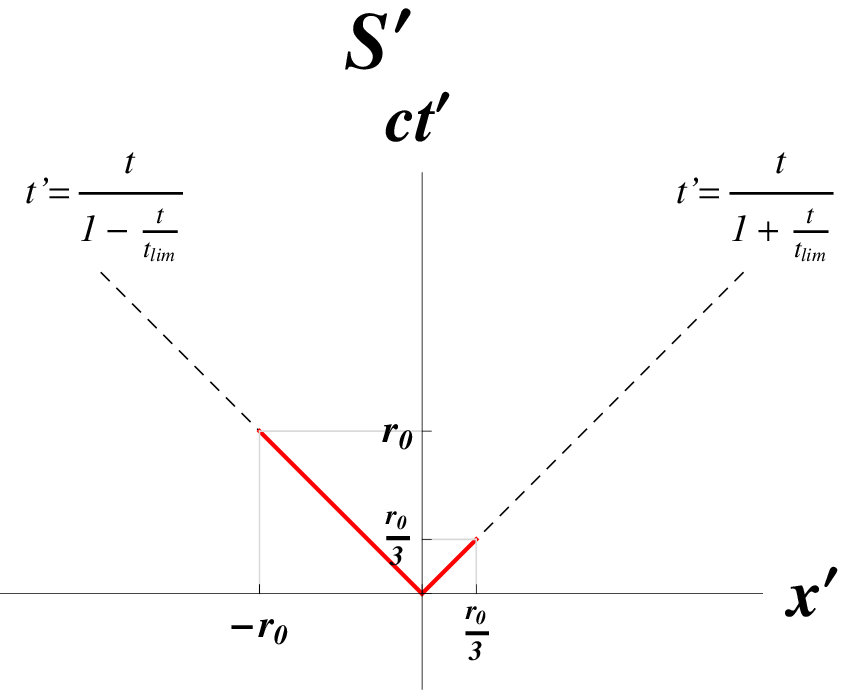} 
\caption{Conformal deformation of the light cone. Initially equal
intervals $t_\pm=\frac{t_{lim}}{2}$ of forward and backward
generatrices transform into unequal intervals
$t'_+=\frac{t_{lim}}{3}$ and $t'_-=t_{lim}$.}
\end{figure}

The non-linear time transformation (12) will further be referred
to as the conformal deformation of time, or conformal time
inhomogeneity.

\section*{III. The dependence of distance on the red shift}
First and foremost we show that transformations (12) allow us to
obtain an explicit expression for location distance as a simple
function of red shift $z$.

Let us consider, on the basis of the formula (12), the case of a
signal propagation from the deep past to the point of the observer
position. That means that we choose the lower sign in the formula
(12): $$t'=\frac{t}{1-\frac{t}{t_{lim}}},\eqno(13)$$ where
$t_{lim}=c/w$, and $t$ is the time of the signal propagation to
the point of observation.

First of all, from (13) we obtain the explicit expression for the
time interval of the signal propagation as a function of the red
shift $z$.

From (13), we have, for small time increments $\Delta t'$ and
$\Delta t$, the following expression
$$
\Delta t'=\Delta t(1-\frac{t}{t_{lim}})^{-2}. \eqno(14)
$$
If $\Delta t$ and $\Delta t'$ are the periods of oscillations of
the emitted ($\Delta t= T_{emitted}$) and received ($\Delta
t'=T_{observable}$) signals, respectively, then, using the
standard definition of the red shift
$$
\frac{\lambda_{observable}}{\lambda_{emitted}}=z+1, \eqno{(15)}
$$
where  $\lambda_{observable}=cT_{observable}$ and
$\lambda_{emitted}=cT_{emitted}$, we find, from (14), the
expression
$$
\frac{\lambda_{observable}}{\lambda_{emitted}}=
(1-\frac{t}{t_{lim}})^{-2}=z+1, \eqno(16)
$$
which gives
$$t(z)=t_{lim}\frac{(z+1)^{1/2}-1}{(z+1)^{1/2}}.\eqno(17)$$
Here $t(z)$ represents the time interval between the moments of
emitting and receiving the light (electromagnetic) signal. So,
assuming that the speed of light is constant and does not depend
on the velocity of the emitter, the quantity $R=ct$ can be
regarded as the distance covered by the signal.

From the formula (17), we obtain an expression which determines
the explicit form of dependence of $R$ on the red shift
$z$:
$$R(z)=R_u
\frac{(z+1)^{1/2}-1}{(z+1)^{1/2}}=R_u\left(1-\frac{1}{(z+1)^{1/2}}\right).
\eqno(18)
$$
Here$R_u=ct_{lim}$ is a parameter, which, within the model
suggested, has the sense of the limit (maximal) distance.

The quantity $R(z)$ defined by (18) corresponds to the distance,
which in cosmology is referred to as a location distance. In
principle, the relation (18) allows for a direct experimental
verification in the whole range of $z$ variation, and can be
confirmed or refused by observations. We are to emphasize the 
essentially kinematic nature of the relation (18). It is the manifestation of the
\underline{nonlinear} conformal time deformation (12) which
follows from Special Conformal Transformations exactly in the same
manner as the Doppler effect, and the dependence $V(z)$ follows
from the \underline{linear} time deformation arising from the
Lorentz boosts leaving the equation of light cone unaltered.
The question of the connection between R(z) and the spectrometric (photometric)
distance adopted in the standard astrophysics calls for special investigation.

\section*{IV. The Hubble law. Connection between the Hubble constant and the
background acceleration}
Now, we can obtain the explicit expression for the Hubble law. Using (18) and
the known expression for the function $V(z)$:
$$
\frac{V(z)}{c}=\frac{(z+1)^2-1}{(z+1)^2+1}. \eqno(19)
$$

We can find the following expression for the ratio $V/R$: 
$$\frac{V(z)}{R(z)}=cR_u^{-1}f(z),\eqno{(20)}$$
where
$$ f(z)=\frac{(z+1)^{1/2}}{(z+1)^2+1}
\cdot\frac{(z+1)^2-1}{(z+1)^{1/2}-1}.$$ It is easy to
see that $\lim_{z\rightarrow0}f(z)=2$.

In this limit we obtain the conventional expression for the Hubble law:
$$V=H_0R,\eqno(21)$$ where $H_0$ is the Hubble constant.

By comparing this formula with $\lim_{z\rightarrow0}
\frac{V(z)}{R(z)}$ from formula (20), we can establish the
following connection between the acceleration $w$ and the Hubble
constant $H_0$: $$2cR_u^{-1}=2w/c=H_0.\eqno(22)$$ It is seen that
the relation (12) defining the conformal time inhomogeneity
allows us to establish the following simple
connection between the parameter $w$ defining the background
acceleration and the Hubble constant $H_0$
$$w=\frac12cH_0.\eqno(23)$$ Hence, in the considered approach the
constant acceleration $w$ intrinsic to non-inertial RF $S'$ can
naturally be connected to the Hubble constant $H_0$, defining
space expansion.

\section*{V. Accelerating Universe effect without Dark Energy}
Now we analyze the general relation (20) for $V(z)/R(z)$.
Taking into account the connection (22) between $R_u$ and the Hubble constant $H_0$, we
rewrite this equation in the dimensionless form as
$$\phi(z)=\frac{V(z)}{H_0R(z)}= \frac12\frac{(z+1)^{1/2}}{(z+1)^2+1}
\cdot\frac{(z+1)^2-1}{(z+1)^{1/2}-1}.\eqno(24)$$

The function $\phi(z)$ 
is shown in Fig. 2. Horizontal line in represents strict Hubble law (21). This
function possesses a maximum at $z_0\cong0.475$.
Overall variation of $\phi(z)$ demonstrates that in the
interval $0\leq z<z_0$ the distance $R(z)$ increases with $z$ more
slowly, and in the interval $z_0<z<\infty$ approaches its limit
value $R_u$ more rapidly, than the velocity $V(z)$ approaches its
limit $c$ (Fig. 3). We see that from the point of view of the proposed approach the
origin of this maximum has the pure kinematic origin.

\begin{figure}
 \begin{center}
 \includegraphics[scale=0.8]{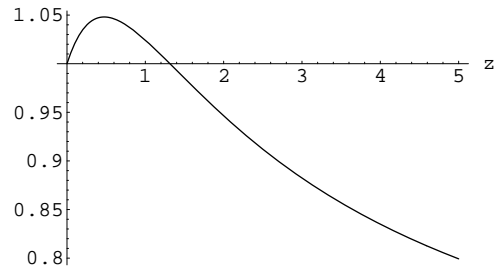}
 \caption{The function $\phi(z)$. }
 \label{Figure5}
 \end{center}
 \end{figure}

 \begin{figure}
 \begin{center}
 \includegraphics[scale=0.8]{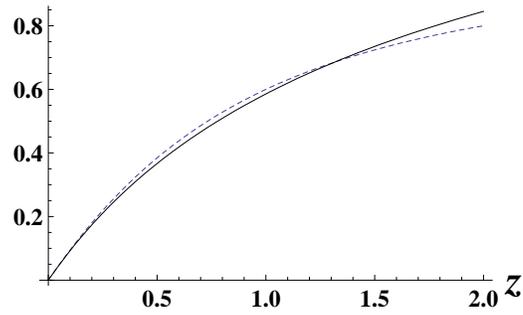}
 \caption{Functions $V(z)/c$ (dashed line) and $R(z)/R_u$ (solid
 line). The values of the both functions coincide at the points of
 $z=0$ and $z\approx1.315$.
 The inequality $V(z)/c>R(z)/R_u$
 ($V(z)/c<R(z)/R_u$) takes place in the interior (exterior) of this
 interval.}
 \label{Figure7}
 \end{center}
 \end{figure}

As regards a possible treatment of the behavior of the function
$\phi(z)$ in terms of the standard dynamical GR approach using the
decelaration parameter, we are to notice the following. According
to the pure kinematic approach proposed in our paper, the source
of the effects induced by the cosmologic expansion is the
conformal time inhomogeneity. The ``acceleration'' attributed to
the emitting source arises because of treating the actual
nonlinear time dependence $t'(t)$ in terms of the traditional
theoretical paradigm based on the time homogeneity concept.

The interpretation of $\phi(z)$ behavior from the point
of view of common treatment seems as follows. In the interval
$\infty > z > z_0$, there is \underline{deceleration}
$\left(\frac{d\phi}{dz}<0\right)$ of cosmological expansion, which
turns to \underline{acceleration}
$\left(\frac{d\phi}{dz}<0\right)$ at $z_0\cong 0{,}475$.
Numerical value of $z_0$ agrees quite well with
experimentally founded ''point of change'' $z_{exp}=0.46\pm0.13$.

It should be emphasized that the basic formula (12) for the
conformal transformations of the time, as well as all its
consequences, are valid on the assumption that the Hubble
parameter $H_0$ is constant. Hopefully this assumption is
reasonable as applied to at least later stages of the Universe
evolution. In this case the proposed formulae (18) and (20) can be
valid for the experimentally  obtained values of the read shift
having the order of several units.

It would be of interest to compare the obtained expression (24) for the function
$V(z)/R(z)$ with such one which can be derived at once from the basic correlation
$$\frac{R_{today}}{R_{emitted}}=\frac{\lambda_{today}}{\lambda_{emitted}}=z+1\eqno{(25)}$$
adopted in the contemporary cosmology. Here $R_{today}$ and $R_{emitted}$ are the Universe
linear sizes at the instant of the reception and the emission of the electromagnetic signal
correspondingly.

Defining the location distance in the euclidean approximation i.~e. neglecting the spatial curvature and using the definition (25) we obtain the condition $R_L(z)=R_{today}\frac{z}{z+1}$, and
taking into account the formula (19) defining $V(z)$ we come to the following expression:
$$\frac{V(z)}{R_L(z)}=cR_{today}\varphi(z),$$
where
$$\varphi(z)=\frac{(z+1)^2-1}{z}\cdot\frac{(z+1)}{(z+1)^2+1}.\eqno(26)$$
The expression above reproduce under the condition $z\ll1$ the stndard form of the Hubble law,
whence it follows that $R_{today}=cH_0^{-1}.$
Thus we obtain the expression for the relation $\frac{V(z)}{H_0R(z)}$, alternative to (24).

It is easy to see that the functions $\phi(z)$ and $\varphi(z)$ coincides in the first approximation on $z\ll1$ ($\varphi(z)|_{z\ll1} \simeq\phi(z)|_{z\ll1}\simeq1+z$) and they possess the maxima. But positions of the maxima and corresponding magnitudes are very different: $z_{max}\simeq1.41$ and $\varphi(z_{max})\simeq1.21$, $z_{max}\simeq0.475$ and $\phi(z_{max})\simeq1.05$ correspondingly. The numerical coincidence of the function $\phi(z)$ with experimental data near the point $z_{exp}=0.46\pm0.13$ can be treated as the strong experimental evidence in favour of the proposed expression $R(z)$ (18) defining the local cosmological distance.

\section*{VI. The universal blue-shifted frequency drift. Pioneer Anomaly}
Now let us consider, on the basis of the formula (12), the
location-type experiments. The conventional scheme of such an
experiment is as follows:
\begin{itemize}
\item[(1)] the signal is emitted from the point of the observer
location at the time instant $t^0_A$,
\item[(2)] the signal is arrived and reemitted at the time instant $t_B$,
\item[(3)] the signal is returned to the observer at the time instant $t_A$.
\end{itemize}

Under the assumption of the coincidence of the forward
$(t_B-t_A^0)$ and backward $(t_A-t_B)$ time intervals, one can
obtain the formula for the signal traveling time
$t=\frac12(t_A-t_A^0)$, and then accept the formula $R=ct$ for the
corresponding location distance.

The time inhomogeneity (12) changes situation such that the
forward and backward time intervals do not coincide. The latter
time interval is larger then the former one (Fig. 4).

\begin{figure}[h!]
\centering \includegraphics[width=0.48\textwidth]{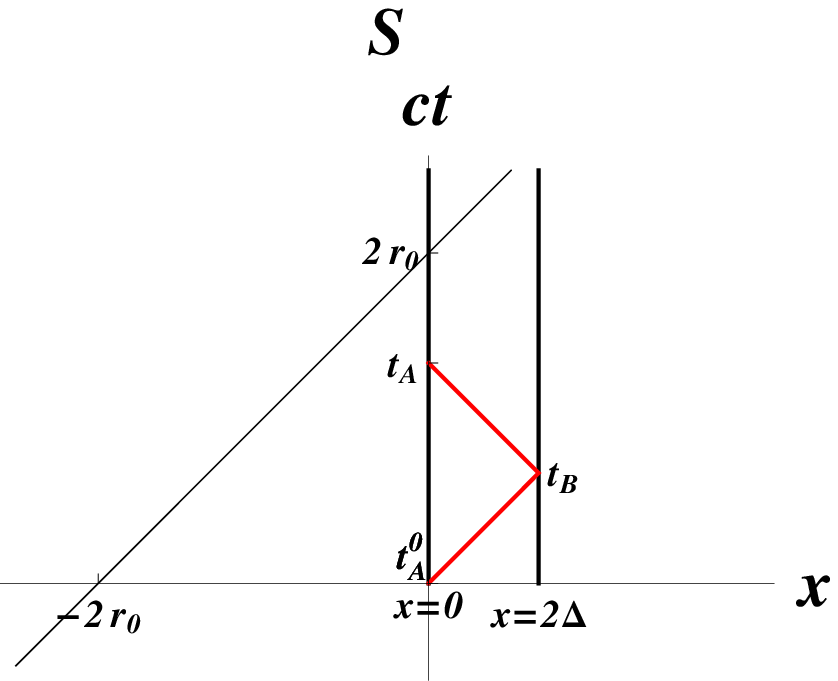}
\includegraphics[width=0.48\textwidth]{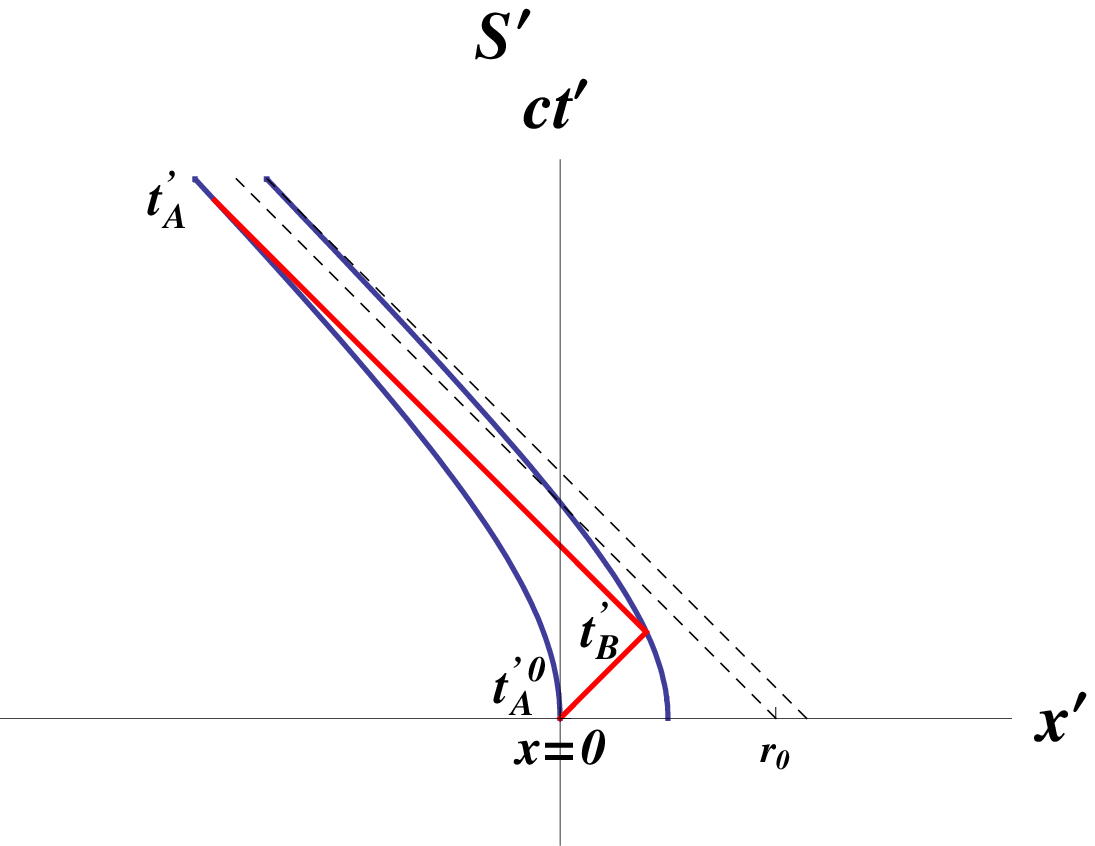} 
\caption{Location-type experiment. World lines of the fixed observer and the object in $S$
transforms into hyperbolas in $S'$. Thin lines are light signal world lines. Under the conformal transformations, light lines remain straight and inclined at $45^\circ$. In RF $S$ we have $t_A-t_B=t_B-t_A^0$, but in $S'$ always the inequality $t'_A-t'_B>t'_B-{t'}_{\!A}^0$ takes
place.}
\end{figure}

In application to real experiment analysis one should use (12) for
the small time intervals, i.e., when $\Delta t/t_{lim} =
(t_A-t_A^0)/t_{lim} \ll 1$. In this case, formula (12) gives, to
the second order of $t/t_{lim}$ $$t'_\pm=t\mp\frac{t^2}{t_{lim}}.
\eqno(26)$$ In accordance with the location distance definition,
the signal travelling distances in forward and backward
directions, respectively, will be
$$x'_\pm=ct'_\pm=x\mp\Delta x=x\mp \frac{W_0}{2}t^2, \eqno(27)$$
where $$W_0=\frac{2c}{t_{lim}}=2w.\eqno(28)$$

Therefore in the $t/t_{lim} \ll 1$ approximation ($t$ is the
signal propagation time) the forward and backward location
distances $x'_\pm$ differ from $R=ct$ by $\Delta
x=\frac{W_0}{2}t^2$. From the usual point of view, it appears that
the emitter fixed in any space point is subjected to constant
acceleration $W_0=2w$ directed to the observer.

Clearly, $t/t_{lim} \ll 1$ condition is equivalent to the
condition $\frac{wt}{2c}\ll1$, $\frac{wx}{2c^2}\ll1$ (see Sec. I),
which formally corresponds to Galilei-Newton kinematics. So, the
predicted effect in the location-type experiments will be the blue
frequency shift, which value will be defined by the formula
similar to (7), i.e.
$$\Delta \nu_{obs}=\nu_0\frac{W_0t}{c}.\eqno(29)$$ Due to (28) the value
of this shift is twice as large as predicted by (7).

For the constant rate of frequency shift
$\dot\nu_{obs}=\frac{d(\Delta \nu_{obs})}{dt}$, we find the
following relation analogous to (8):
$$\dot\nu_{obs}=\frac1c\nu_0W_0. \eqno(43)$$ Since by (23) $W_0=cH_0$,
where $H_0$ is the Hubble constant, from (30) we have
$$\dot\nu_{obs}=\nu_0H_0. \eqno(31)$$ This relation defines the
frequency drift as a function of the fixed emitter frequency
$\nu_0$.

According to the approach under consideration the anomalous
blue-shifted drift is the consequence of the non-inertiality of the observer's RF.
It can be observed in principle under suitable conditions (in the
absence of any gravitating sources) on any frequency even in the
case of mutually fixed emitter and receiver (see \cite{9,10,11,12}). From this point of
view PA should be treated as the first clearly observed effect of
that kind. The uniform blue-shifted drift $(\dot\nu_{obs})_P$ is measured experimentally with a great
accuracy $(\dot\nu_{obs})_P=(5.99\pm0.01)\cdot10^{-9}$ Hz/s
\cite{3,4}. Therefore it can form a basis for the new (alternative to
the cosmological observations) high precision experimental
estimation of the numerical value of the Hubble constant.

For that goal, we make use of (31), and recall that frequency of
Pioneer tracking is $(\nu_0)_P=2.29\cdot10^{9}$ Hz, such that
$$H_0=\frac{(\dot\nu_{obs})_P}{(\nu_{obs})_P}
\cong2.62\cdot10^{-18}\ \mbox{s}^{-1},\eqno(32)$$ what is
consistent with generally accepted value of
$H_0\cong2.4\cdot10^{-18}\ \mbox{s}^{-1}$ obtained from
cosmological observations.

For the ''acceleration'' $a_P$ we have $a_P=cH_0=7.85\cdot10^{-8}\ \mbox{cm/s}^2,$ what is in the range
of uncertainty of PA data ($a_P^{exp}=(8.74\pm1.33)\cdot10^{-8}$ cm/s$^2$).

The numerical coincidence of the results can be considered as
experimental evidence of anomalous blue-shifted drift as a
kinematical manifestation of the conformal time inhomogeneity. In
other words, from the view point of the considered approach, the
quantities measured in experiments of electromagnetic wave
propagation favor relation (29) (but not (7)) for the anomalous
frequency shift.

It should be stressed that the physical meaning of the relations
(7) and (29) is fundamentally different in spite of their visual
similarity.

Equation (7), defining blue frequency shift by the
non-inertiality of RF $S'$, in fact was obtained in the
Galilei-Newton kinematics. There time transformation under
transition from RF $S$ to $S'$ has the form $t=t'$ (see (4)).

On the other side, formula (29) was obtained from the exact
non-linear time transformation (12) defining the time
inhomogeneity, that is, beyond the Galilei-Newton kinematics.
``Constant acceleration'' $W_0=2w$ appears due to the quadratic
character of the first non-linear term in the power series
expansion of $t'(t)$ in terms of small parameter $t/t_{lim}$ in
(12), while the location distance is defined as $R'=ct'$.
Hence the ``acceleration'' $W_0$ is not a ``truly'' acceleration
(i.e. its origin is not a force or a dynamical source) but rather
a ``mimic'' acceleration.

The possibility of assigning the anomalous frequency drift
observed in the signals transmitted by Pioneer 10/11 to the
quadratic time inhomogeneity was noted in the first comprehensive
works on PA \cite{3} and \cite{4}. However no theoretical reasons were
mentioned for such a phenomenological approach.

On the other hand, the acceleration parameter $w=\frac12cH_0$ seems to be a strictly natural candidate for the ``minimal acceleration'', which is a fundamental dynamical parameter of the Modified Newton Dynamics (MOND) (see \cite{13,14}), which is a phenomenological alternative for the Cold Dark Matter approach.

\section*{Conclusions}
The conformal time inhomogeneity leads to the following consequences:
\begin{itemize}
\item[---] The cosmological location distance can be determined as an explicit 
function of red shift $R(z)$. The combination of this function with the 
SR expression for the longitudinal Doppler-effect $V(z)$ gives the explicit 
analytic expression for the ratio $V(z)/H_0 R(z)$. The ratio coincides 
with the Hubble law in the limit of $z\ll1$, and possesses a maximum at 
$z_{max}\simeq0.475$. The appearance of this maximum is a pure kinematic 
manifestation of the time inhomogeneity and does not need any special
gravitating sources (like the dark energy).

\item[---] In the location-type experiments uniform blue-shifted frequency drift appears, 
which mimics constant acceleration $w=cH_0$ directed towards the observer. The Pioneer Anomaly 
is the first really observed effect of that kind. The observed drift can be used 
for local experimental determination of Hubble constant. This effect can be observed 
in principle at any frequency even between mutually moveless emitter and 
receiver in the absence of any gravitating centers.
\end{itemize}

The topics discussed in the present report was considered in part in author's papers \cite{7,8,9,10,11,12} as well.

\end{document}